# Characterizing Research Leadership on Geographical Weighted Collaboration Network[1]

Chaocheng He[a,b], Jiang Wu[a*], Qingpeng Zhang[b*]

a. School of Information Management, Wuhan University, Wuhan, Hubei, China.
b. School of Data Science, City University of Hong Kong, Kowloon, Hong Kong, China.

**Abstract**

Research collaborations, especially long-distance and cross-border collaborations, have become increasingly prevalent worldwide. Recent studies highlighted the significant role of research leadership in collaborations. However, existing measures of the research leadership do not take into account the intensity of leadership in the co-authorship network. More importantly, the spatial features, which influence the collaboration patterns and research outcomes, have not been incorporated in measuring the research leadership. To fill the gap, we construct an institution-level weighted co-authorship network that has two types of weight on the edges: the intensity of collaborations and the spatial score (the geographic distance adjusted by the cross-border nature). Based on this network, we propose a novel metric, namely the spatial research leadership rank (SpatialLeaderRank), to identify the leading institutions while considering both the collaboration intensity and the spatial features. Harnessing a dataset of 323,146 journal publications in pharmaceutical sciences during 2010-2018, we perform a comprehensive analysis of the geographical distribution and dynamic patterns of research leadership flows at the institution level. The results demonstrate that the SpatialLeaderRank outperforms baseline metrics in predicting the scholarly impact of institutions. And the result remains robust in the field of Information Science & Library Science.

# 1. Introduction

In the "Big Science" era, research collaboration is playing an important role in knowledge creation, iteration, and dissemination (Katz and Martin 1997, Gazni, Sugimoto et al. 2012). Thanks to the increasing scale and complexity of scientific projects, we have witnessed a rapid growth in the frequency and influence of collaborative research (Wang, Xu et al. 2013, Chen, Zhang et al. 2019). This trend could be evidenced by the increase in research collaborations at a distance (Hoekman, Frenken et al. 2010, Bercovitz and Feldman 2011).

It is important to gain an in-depth understanding of the trend towards collaborations at a longer distance and even cross-border for the following reasons. First, it enhances the quality of research by combining the expertise and resources and sharing the cost (Hoekman, Frenken et al. 2010). Compared with single-authored publications, collaborative work often leads to more novel and higher internal quality control (Fernandez, Ferrandiz et al. 2016). And these benefits increase as the distances between collaborators increases, as potential collaboration partners are more likely to be found (Hoekman, Frenken et al. 2010). Indeed, cross-border collaborative research tends to be of higher quality and greater impact compared with that of local collaborators (Guan, Zuo et al. 2016, Jiang, Zhu et al. 2018). Second, because of the aforementioned benefits, significant public policies and expenditures have been established and allocated to facilitate long-distance research collaborations (Hoekman, Frenken et al. 2010). For example, China's 12th Five Year (2011-2015) Plan for Science and Technology Development stated that China will actively take part in international science and technology organizations and cross-border research collaborations (Wang and Wang 2017). Similarly, the European government attempts to construct a *European Research Area* by coordinating regional, national and EU research activities (Hoekman, Frenken et al. 2010) to improve European states' internal consistency and to break down barriers to innovate. Third, many countries and regions are making great efforts to attract overseas talents, who further facilitate the cross-border collaborations (Baruffaldi and Landoni 2012). For example, the 1000-talent program of China has attracted more than 7000 full-time and part-time senior scholars by 2018. Many are the leader in the field and 22.5% are non-Chinese researchers (Jia 2018). Within the first five years of the 1000-talent program, China has attracted more full professors than the past 30 years combined[2]. These talents coming from overseas bridged the gap between the research communities in two countries and greatly improved both the intensity and quality of cross-border collaborations.

Existing literature on spatial research collaboration has two main limitations. First, most studies assume that the research collaborations are homogeneous while ignoring the existence of leadership in the collaboration. It is sensible that the first and corresponding author(s) play a leading role in the collaboration, and thus their relationship with other authors should be stronger than those among others (Wang, Wu et al. 2014). There are several studies examining the research leadership (Wang, Wu et al. 2014, González-Alcaide, Park et al. 2017). However, they either do not consider the social network structure of the co-authorship network (Chaocheng, Jiang et al. 2019) or ignores the intensity of collaborations while measuring the leadership (Zhou, Zeng et al. 2018). Second, collaboration relationship is assumed to be evenly distributed to authors of the same article, without considering distances between the institutions. Existing algorithms for identifying key research entities in collaborations mainly focus on the topological features of the entities in the collaboration network (Zhou, Zeng et al. 2018), without incorporating the spatial features into the network. Collaboration at a longer distance has a higher impact than that at a short distance (Hoekman, Frenken et al. 2010). Identifying important research entities in research collaboration from a spatial perspective becomes necessary and essential to knowledge creation as well as dissemination (Forman and van Zeebroeck 2019) (Wu 2013, Copiello 2019).

To fill the above research gaps, we extend the literature as follows. First, we model the collaboration relationships as a directed network at the institution level, where the direction of an edge indicates the leadership flow between the two institutions. Second, we incorporate the geographical distance between institutions as a weight on the edge in the collaboration network. Third, based on the constructed geographical weighted collaboration network, we propose a novel metric, namely the spatial research leadership rank (SpatialLeaderRank), to identify the leading institutions while considering both the collaboration intensity and the spatial features. More specifically, an institution is considered with higher leadership status according to the following three criteria: (a) the institution frequently plays the corresponding rule in papers with other institutions; (b) the institution frequently plays the corresponding rule in longer distance and even cross-border collaborations; (c) the participating institutions have high leadership status themselves (i.e. it is leading other leaders). We exemplify and validate the proposed SpatialLeaderRank metric using the journal publications in the pharmaceutical sciences, a field that has witnessed a dramatic increase in collaborations between multinational scientists in both academia and biotechnology sector because of the need for very diverse expertise from various disciplines and the rise of R&D outsourcing (Herrling 1998, McKelvey, Alm et al. 2003, Plotnikova and Rake 2014). We also examine the SpatialLeaderRank in the Information Science & Library Science as the robustness check.

---

[2] http://kfq.job1001.com/

In the following sections, we review the literature in Section 2, and then describe the data and methods in Section 3. The results are presented in Section 4. We conclude the paper with discussions of the limitations and the implications for theory and policymaking in Section 5.

## 2. Literature review

### 2.1. Spatial research collaboration

The existing research on the spatial research collaboration mainly focuses on the spatial pattern of research collaboration and the role of geographical proximity on research collaboration. The spatial pattern of research collaboration has been systematically studied at multiple levels. A recent study illustrated the spatial patterns of cross-border knowledge flows and evaluates the effect of various factors including the geographical factor (Gui, Liu et al. 2018). The establishment of research alliances with more developed countries constituted a critical mechanism, which integrated developing countries into the global research community (González-Alcaide, Park et al. 2017). Researchers in small European states were found to be less homogenously collaborating with both domestic and foreign partners (Ukrainski, Masso et al. 2014). Research activities in Brazil were found to be spatially heterogeneous and a geographical decentralization process of scientific research activities across the country was needed to stimulate the development of those privileged areas (Sidone, Haddad et al. 2017).

The role of geographical proximity has been explored in various fields. Plotnikova and Rake (2014) examined the country-level determinants in pharmaceutical research and found that the geographic distance was negatively associated with the cross-border research collaboration. In humanities, arts and social sciences, the geographical distance was found to be critical to the collaborative activities (Luo, Xia et al. 2018). Similar conclusions could be drawn from the field of ecology (Parreira, Machado et al. 2017), and immunology (Lander 2015), and so on.

Research collaboration at a distance has become increasingly prevalent (Hoekman, Frenken et al. 2010, Bercovitz and Feldman 2011). In all countries and all academic fields, cross-border research collaboration tends to result in higher publishing rates and impact (Kwiek 2015). Researchers who collaborate at a distance are often better performing researchers (Jonsen, Butler et al. 2013, Wagner, Whetsell et al. 2019). Given the abundant evidence that the spatial features influence the research performance, existing research on collaboration networks did not account for the spatial patterns, and mainly assumed that the edges are homogeneous in the collaboration network.

## 2.2. Research leadership

It is commonly recognized that in the same publication, the first author and the corresponding author often lead the research collaboration and make a major contribution (Wang and Wang 2017). Typical instances include the field of biology and engineering. In the biomedical field, the first author is often an early-career researcher who is assigned to carry out the research and write the research paper. Simultaneously, other non-leading (participating) coauthors act more specialized roles, such as contributing statistical analyses or data visualization (Sekara, Deville et al. 2018). However, corresponding authors should be responsible for both scientific and non-scientific contributions such as designing the roadmap of a project, assigning research tasks, guiding the experiments, and checking the logic of the paper, for coordinating the completion and submission of the work (Wang and Wang 2017). The corresponding author's responsibility is more prominent with the increase of collaboration scales, the growing complexity and depth of the research (Hemlin, Allwood et al. 2013). In most cases, the first and corresponding authors belong to the same organization (Wang, Wu et al. 2014). Hence, the corresponding author's affiliation is considered the research leader of a research paper in this study. Chaocheng, Jiang et al. (2019) define that each paper possesses a total leadership mass of 1, and thus the intensity of the research leadership flow from the leading institution $a$ to institution $b$ in the paper $i$, $RL_{ab,i}$ is obtained as

$$RL_{ab,i} = \frac{1}{LIN_i} \times \frac{1}{N_i}, \tag{1}$$

where $LIN_i$ is the number of leading institutions and $N_i$ is the number of institutions in paper $i$. The aggregated intensity of RL flow from institution a to institution $b$ in all their co-publications, $C_{ab}$, is defined as follows,

$$C_{ab} = \sum_{i=1}^{M_{ab}} C_{ab,i}, \tag{2}$$

where $M_{ab}$ is the number of papers where $a$ is the leading institution and $b$ is a participating institution. Institution $a$'s total research leadership mass (the number of publications where institution $a$ is the leading institution), $LM_a$, is defined as follows,

$$LM_a = \sum_{b=1}^{B} C_{ab}, \tag{3}$$

where $B$ is the number of institutions that $a$ has led. We refer the readers to (Chaocheng, Jiang et al. 2019) for details of the research leadership flow and research leadership mass.

## 2.3. PageRank and LeaderRank

PageRank was originally proposed by Google to rank the importance of webpages (Brin and Page 1998). There is a boom in its variation and application in a broad set of fields in the following decades. PageRank has been widely applied to analyzing the research collaboration network. Liu, Bollen et al. (2005) transformed each undirected edge in the research collaboration network into a set of bi-directional, symmetrical edges, and defined modification of PageRank, namely the AuthorRank. Fiala, Rousselot et al. (2008) modified the PageRank by incorporating both citation and co-authorship graph property.

Successful as it is, PageRank has several drawbacks (Gleich 2015). The stability of ranking and the robustness to noise and manipulation vary given different parameters (Lu, Chen et al. 2016). Moreover, if there are disconnected components in the network, the ranking result is not unique. To this end, Lu, Zhang et al. (2011) proposed the LeaderRank, an adaptive and nonparametric algorithm, by adding a ground node that bi-directionally connected to every other node and then performing random walks. As a result, LeaderRank has a faster convergence rate, higher stability for noisy data, and more robustness to manipulations (Li, Zhou et al. 2014). LeaderRank is further applied to identify the influential nodes in complex products and systems (Li, Wang et al. 2019); in power grids (Zhou, Lei et al. 2019); in manufacturing services (Wu, Peng et al. 2019). However, it is noteworthy that, all the previous variants of PageRank including LeaderRank only consider the topological features of nodes, while ignoring other non-topological features, especially, the spatial features that is found to be very important to the academic performance and impact.

## 3. Data and methodology

### 3.1. Data collection

We leverage the Web of Science Core Citation Database and perform a data collection to construct the geographical weighted and directed network. Specifically, following (Plotnikova and Rake 2014) we collect publications in categories ( "Biochemistry and Molecular Biology", "Biotechnology and Applied Microbiology'', "Chemistry, Applied", "Chemistry, Medicinal", "Medicine, Research and Experimental", "Pharmacology and Pharmacy" and "Toxicology") related to pharmaceutical study during 2010-2018. To check the robustness and have a comprehensive understanding, we also collect the publications in the category of "Information Science & Library Science (ISLS)" and compare the key results in these two very different fields. More precisely, the data was retrieved using the search term "WC = A AND PY = B", where A is the above sub-categories and B it 2010-2018. We restrict publications of journal articles and exclude other non-journal publications such as meeting abstracts, letters, editorial materials or reviews. We further restrict the data and sample 323,146 publications in pharmaceutical sciences and 28,158 publications in ISLS that have at least two co-institutions. To construct the weighted and directed spatial research leadership network, we extract 2459 institutions in pharmaceutical sciences and 841 institutions in ISLS, which have been the primary affiliation of the corresponding author for at least one paper (with multiple institutions) in each year. We utilize Google Map to obtain the latitude, longitude, and the country of each institution.

## 3.2. Geographical weighted and directed network

Spatial distance is one of the classic proximities for research collaborations (Boschma 2005). There are plenty of evidence that the spatial distance is a hinder determinant to the formation of research collaborations (Fernandez, Ferrandiz et al. 2016, Zhang and Guo 2017). Recently, cross-border collaborations are becoming increasingly popular. Cross-border collaboration brings unique benefits (Jiang, Zhu et al. 2018), including the better access to international data (Jonsen, Butler et al. 2013), the higher tendency to stimulate new ideas (Ellis and Zhan 2011), and increased international visibility and impact (Kwiek 2015). Therefore, in this study, we propose to measure the spatial proximity using the following spatial score, which takes an additive form of both the distance and the cross-border nature. For a publication $i$, the spatial score of leading institution $a$ and participating institution $b$ pair is defined as:

$$SPS_{i,ba} = lg(Distance_{ba}) + \lambda Country_{ba}, \qquad (4)$$

where $Distance_{ba}$ denotes the geographical distance between institutions $b$ and $a$. $Country_{ba}$ is a dummy variable indicating if $b$ and $a$ are from different countries. $\lambda$ represents the relative importance of the cross-border nature of the collaboration. $\lambda$ is a field-specific parameter because of the varying occurrence of the cross-border collaboration for different research fields. The value of $\lambda$ is obtained by calculating the ratio between the coefficient of $Country_{ba}$ and the coefficient of $Distance_{ba}$ in the gravity model (Fernandez, Ferrandiz et al. 2016, Zhang and Guo 2017), as described by the following equation:

$$I_{ab} = \beta_0 + \beta_1 lg\, Pubmass_a + \beta_2 lg\, Pubmass_b + \beta_3 lg\, Distance_{ab} + \beta_4 Country_{ab} + \sum_{k=5}^{K} \beta_k lgs_k + \epsilon_{ab}, \qquad (5)$$

where $I_{ab}$ denotes the collaboration intensity, measured by the number of co-publications, $Pubmass_a$ and $Pubmass_b$ denote the number of previous publications of institutes $a$ and $b$, respectively. $s_k$ denotes other dimensions of proximity, including cognitive proximity, social proximity and economic proximity. The gravity model results showed that, during 2010-2018, $\lambda$ was 1.32 and 3.23 in pharmaceutical sciences and ISLS, respectively, indicating that the hindering effect of cross-border nature in pharmaceutical sciences was lower than that in ISLS. We can also obtain the evolution of $\lambda$ by running the gravity model for each year. Figure 1 shows the yearly evolution of $\lambda$ during 2010-2018. The value of $\lambda$ for pharmaceutical sciences was stable, with a lightly decreasing trend, indicating that the hindering effect of cross-border nature has been declining over time. While the value of $\lambda$ for ISLS oscillated (ranging between 2.76 and 3.77).

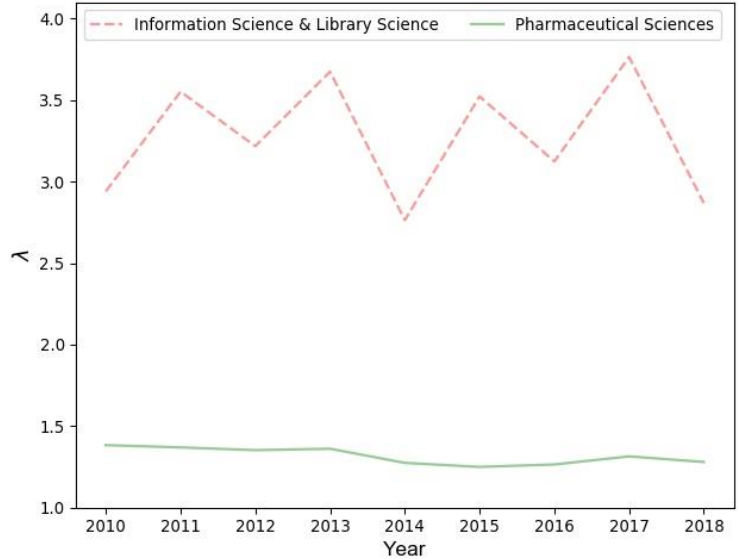

Figure 1 The values of $\lambda$ for pharmaceutical sciences and ISLS during 2010-2018.

Therefore, the longer the geographical distance is, the smaller the probability of collaboration and the higher the spatial score are. Meanwhile, the cross-border nature positively influences the spatial score. The spatial score of a publication $i$ is the average spatial score of the leading institution and participating institution pairs, which can be expressed as

$$SPS_i = \frac{1}{N_a} \times \frac{1}{N_b} \times \sum_{a}^{N_a} \sum_{b}^{N_b} SPS_{i,ba}, \qquad (6)$$

where $N_b$ is the number of institutions and $N_a$ is the number of leading institutions (i.e. more than one correspondence institutions). For a publication $i$, we define the spatial research leadership flow intensity from participating institution $b$ to leading institution $a$ as follows[3],

---

[3] In (Chaocheng, Jiang et al. 2019), the direction of research leadership is from the leading institution to the participating institution. Differently, to better depict the process of gaining the leadership from leading others, the direction is defined to be from the participating institution to the leading institution. This definition also makes the following calculation more intuitive and consistent with the network science literature.

$$SL_{i,ba} = SPS_i \times \frac{1}{N_a} \times \frac{1}{N_b}, \tag{7}$$

The total spatial research leadership flow intensity from $b$ to $a$ is calculated as

$$SL_{ba} = \sum_{i=1}^{N_{ba}} SL_{i,ba}, \tag{8}$$

where $N_{ba}$ is the number of papers where $a$ is the leading institution and $b$ is a participating institution.

Now, we can construct a geographical weighted and directed network where institutions are nodes and research leadership flows are directed edges. The weight on the edge represents the spatial research leadership flows' intensities between the two institutions. Figure 2 illustrates the construction of the network.

| Publication 1 | Publication 2 | Publication 3 |
|---|---|---|
| Author 1, Institution *a*, Country *A*<br>Author 2, Institution *b*, Country *A*<br>Author 3, Institution *c*, Country *B*<br>Author 4, Institution *a*, Country *A*<br>Author 5, Institution *d*, Country *C* | Author 3, Institution *c*, Country *B*<br>Author 4, Institution *a*, Country *A*<br>Author 5, Institution *d*, Country *C*<br>Author 2, Institution *b*, Country *A* | Author 1, Institution *a*, Country *A*<br>Author 3, Institution *c*, Country *B* |
| **Corresponding Author**<br>Author 4, Institution *a*, Country *A* | **Corresponding Author**<br>Author 3, Institution *c*, Country *B* | **Corresponding Author**<br>Author 1, Institution *a*, Country *A* |

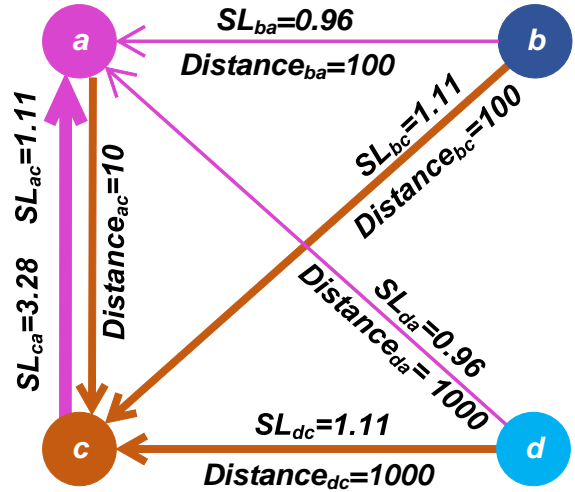

Figure 2. An illustration of a hypothetical geographical weighted and directed network.

## 3.3. Spatial research leadership rank

LeaderRank is a simple variant of PageRank but has been widely proved to outperform PageRank regarding ranking effectiveness with good robustness (Li, Zhou et al. 2014). In LeaderRank, a ground node, which bi-directionally connects to every other node, is added to the existing network with $N$ nodes and $M$ weighted directed edges. Figure 3 is an illustration of adding the ground node to the weighted and directed spatial research leadership network. Thus, the network consists of $N + 1$ nodes and $M + 2N$ edges, and forms a strongly connected network. LeaderRank performs a standard random walk process to rank every node. Successful as it is, LeaderRank only considers the topological features of the collaboration network, while ignoring other non-topological features, especially, the spatial features, which have been recognized as important factors of the academic performance and research impact. Therefore, we propose a new metric, namely the SpatialLeaderRank, to incorporate the spatial features into the measure of research leadership. Specifically, the SpatialLeaderRank of node $a$ at the time step $t$ denoted as $SpatialLeaderRank_a(t)$. Thus, the dynamics of SpatialLeaderRank is described by the following iterative process,

$$SpatialLeaderRank_a(t+1) = \sum_{b=1, b \neq a}^{N+1} \left[ \frac{SL_{ba}}{\sum_{c=1, c \neq b}^{N+1} SL_{bc}} \times SpatialLeaderRank_b(t) \right], \quad (9)$$

where $SL_{ba}$ is the spatial research leadership flow intensity from institution $b$ to $a$. If there is no research leadership from $b$ to $c$, $SL_{bc} = 0$. The spatial research leadership flow intensity from other institutions to the ground node and from the ground node to other institutions is set to 1 (Lu, Zhang et al. 2011). The process starts with the initialization where all institutions' SpatialLeaderRank being equal to 1. According to the iterative process described by equation (9), the SpatialLeaderRank value will converge to a unique and steady-state $SpatialLeaderRank_a(\infty)$, ($a = 1,2,...,N, N + 1$). We rank all institutions according to $SpatialLeaderRank_a(\infty)$.

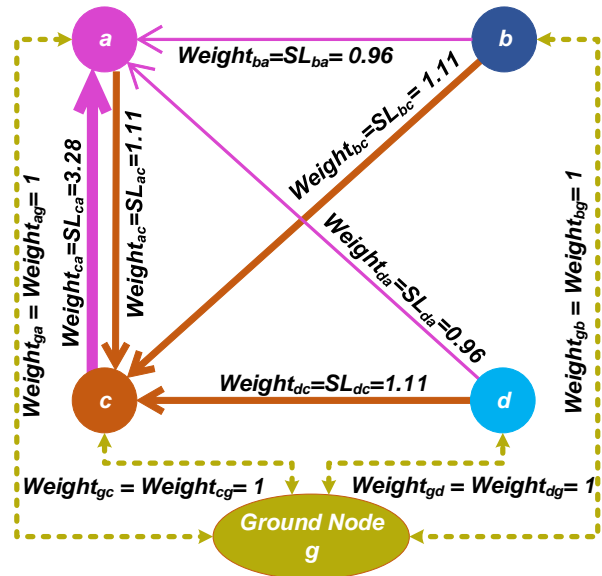

Figure 3 An illustration of adding the ground node to the geographical weighted and directed network to calculate SpatialLeaderRank.

## 4. Result and analysis

For simplicity, we mainly introduce the results for pharmaceutical sciences, and then summarize the results for ISLS in sub-section 4.2.4.

## 4.1. Spatial pattern and evolution of research leadership flows in pharmaceutical sciences

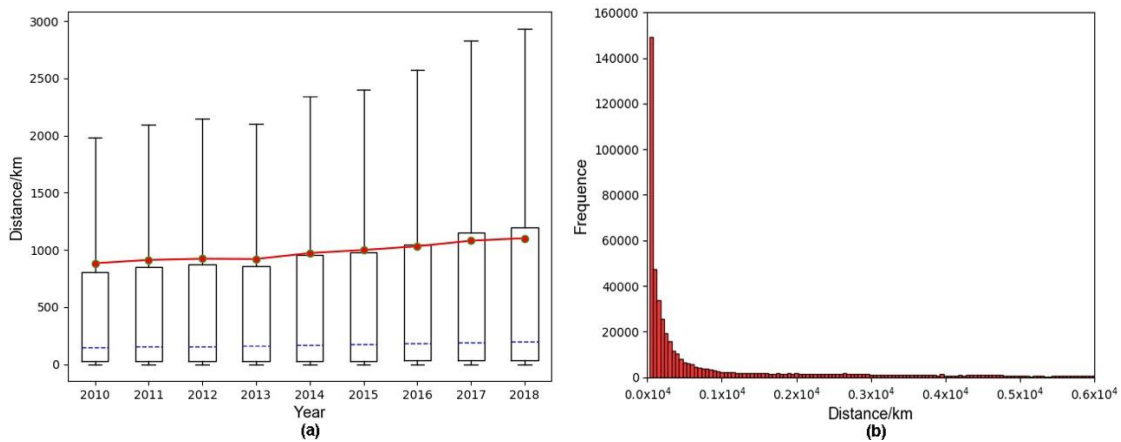

Figure 4. (a) The yearly boxplot of the distance of research leadership flow during 2010-2018. (b) The overall distribution of the distance of research leadership flow during 2010-2018. Red dots denote the data. The green dashed line is the fit of the power-law distribution.

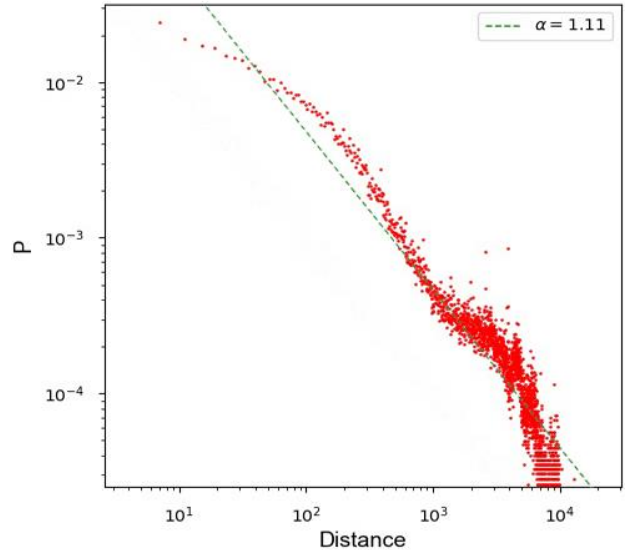

Figure 5 The power-law distribution of the distance of research leadership flow during 2010-2018.

Figure 4 (a) presents the yearly boxplot of the distance of research leadership flow during 2010-2018, with the blue dotted line representing the median and red dots representing the mean. The increasing trend of both the median and mean indicate that, in general, the research leadership flow distance has been increasing steadily over time. Institutions are leading or being led by institutions located far away with an increasing frequency. Similarly, the distribution of research leadership flows exhibits a long-tail distribution (see Figure 4 (b)), with most research leadership flows occurring among institutions that are geographically close to each other, and in the meantime, the distance of research leadership flows can be very long (cross continent) for a small amount of cases. The data can well fit a power-law distribution $P(k) \propto k^{-\alpha}$, where $k$ is the distance and $\alpha$ is the exponential parameter as shown in Figure 5. The value of $\alpha$ was decreasing over time (from 1.13 in 2010 to 1.10 in 2018), indicating that the decaying speed was increasing. To sum, the long-distance research leadership flow has become increasingly popular during 2010-2018.

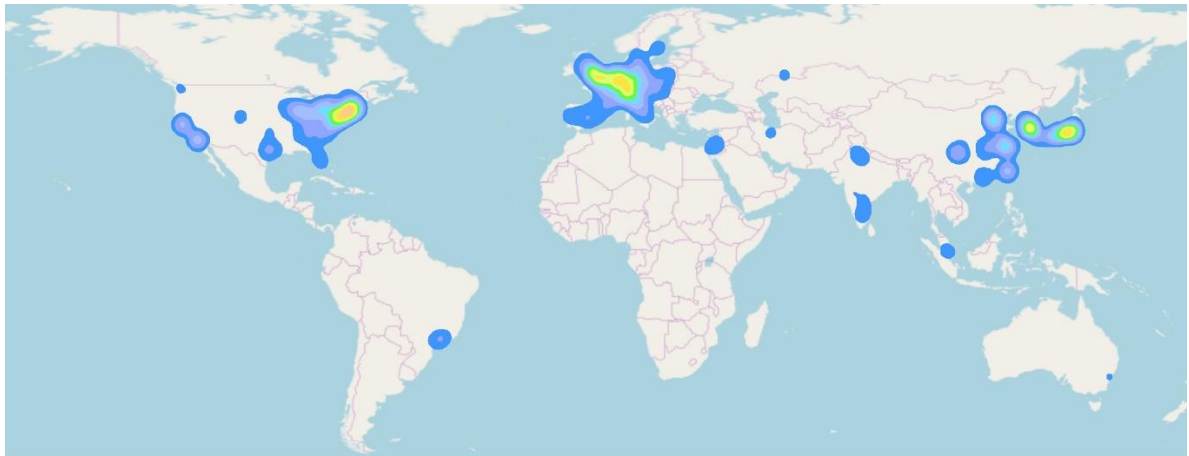

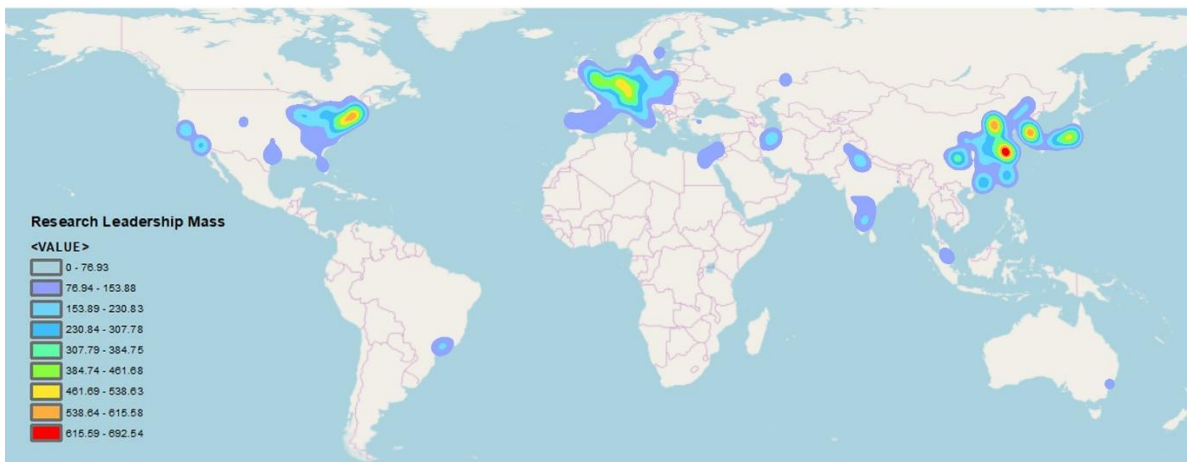

Figure 6. The kernel density heat map of the distribution of research leadership mass at the regional level.

To present the change in the research leadership mass in pharmaceutical sciences, we visualize the geographical distribution of research leadership mass in Figure 6, the kernel density heat map of global research leadership mass in two split time periods, 2010-2014 and 2015-2018. The kernel density estimation smooths the spatial coordinates to generate a probability density surface of a set of point locations. For the detailed calculation of the kernel density estimation, please refer to (Downs and Horner 2012). We can identify three main research leadership mass clusters, Northeastern United States, European Union, and Northeastern Asia. More specifically, during 2010-2014, the cluster of European Union covered the largest area and the Northeastern United States had the highest density. In the Northeastern Asia cluster, Japan and Republic of

Korea led the region. In the Greater China Region, the research leadership mass was mainly distributed in the eastern part. In general, during 2010-2014, the pharmaceutical sciences research was dominated by the most economically developed countries. Differently, during 2015-2018, although these developed countries were still playing the prominent role in leading the pharmaceutical research, Eastern Asia has emerged as a key leader, with multiple significant clusters in China, Japan and Republic of Korea. In particular, the Yangtze River Delta Region (the densest region in the figure, including Shanghai, Jiangsu, Anhui and Zhejiang Provinces), the Jing-Jin-Ji Metropolitan Region (including Beijing, Tianjin and Hebei Provinces), and Sichuan Province have emerged as prominent leaders in pharmaceutical research, because the finest universities and medical schools locate in these regions. Other developing countries also had great improvement, especially India and Iran. In general, the distribution of research leadership mass has become more balanced between western and eastern countries.

Figure 7 intuitively presents the detailed evolution of research leadership mass over year with a bump graph, where the width of the bump is proportional to the research leadership mass of the corresponding country or region. We present the top 15 countries/ regions in terms of their research leadership mass. At the first glance, the bump lines of all countries/regions were widening, indicating that the research leadership masses of all were growing during 2010-2018. In line with Figure 6, some significant changes were underway. Mainland China has taken the #1 from the United States in 2018. Japan retained the place at #3. India and Iran have risen to #4 and #5, respectively. In a nutshell, given these changes during 2010-2018, it is apparent that the distribution of research leadership mass is rebalancing between western and eastern countries.

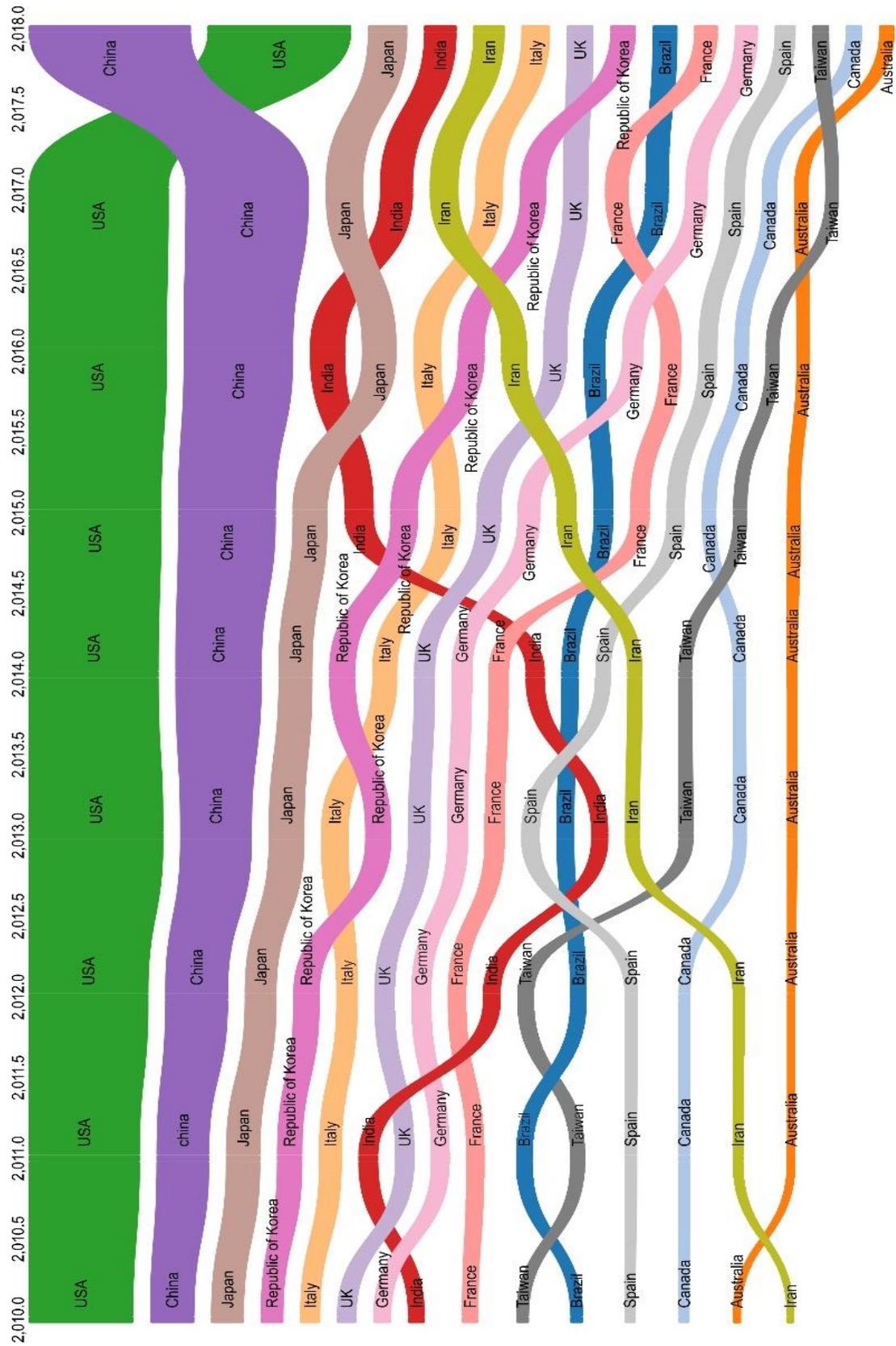

Figure 7. The bump graph of research leadership mass over the year at the country level.

## 4.2. Effectiveness analysis of SpatialLeaderRank

To evaluate the effectiveness of the proposed SpatialLeaderRank, we examine the correlation between the SpatialLeaderRank and other conventional indices (indices of collaboration network and publication number). We then perform the receiver operating characteristic (ROC) curve analysis to evaluate the ability of SpatialLeaderRank and other conventional indices to identify the top 5% institutions with high academic impact. We further use *Ksim* (Haveliwala 2003) to measure the similarity between each indices' ranking results and the academic impact rank. Last, we also compare the detailed rank of institutions in terms of SpatialLeaderRank, other conventional indices, and academic impact indices. The conventional indices include PageRank, betweenness centrality, closeness centrality, indegree centrality, and publication number (Wu 2013, Kim and Diesner 2015). And the academic impact indices include citation count, citation-based h-index, altmetrics count, altmetrics-based h-index.

### 4.2.1. Correlation between SpatialLeaderRank and other conventional indices

We present the scatter plot between the SpatialLeaderRank and other conventional indices in Figure 8, and the Spearman's correlation coefficients in Table 1. We find that all pairs of indices are positively correlated. In addition, except for the pair of closeness and other indices, all other pairs of indices are linearly correlated. More specifically, the Spearman's correlation coefficients for all pairs of indices are larger than 0.833 ($p \leq 0.001$), indicating that all the indices are highly positively correlated with each other. The correlation between the SpatialLeaderRank and LeaderRank is 0.992, indicating that the rank of institutions is identical for the majority of institutions. This is expected because most research collaborations took place among local institutions (see Figure 4b). The difference is mainly among those highly ranked institutions, and cannot be well captured by the correlation coefficient. We will find the difference between them in Section 4.2.2. Another intuitive observation is that the number of publications has the second-highest correlation with SpatialLeaderRank. It is worth noting that the actual causality cannot be revealed by the correlation analysis. It is interesting to explore the causality between these indices in the future.

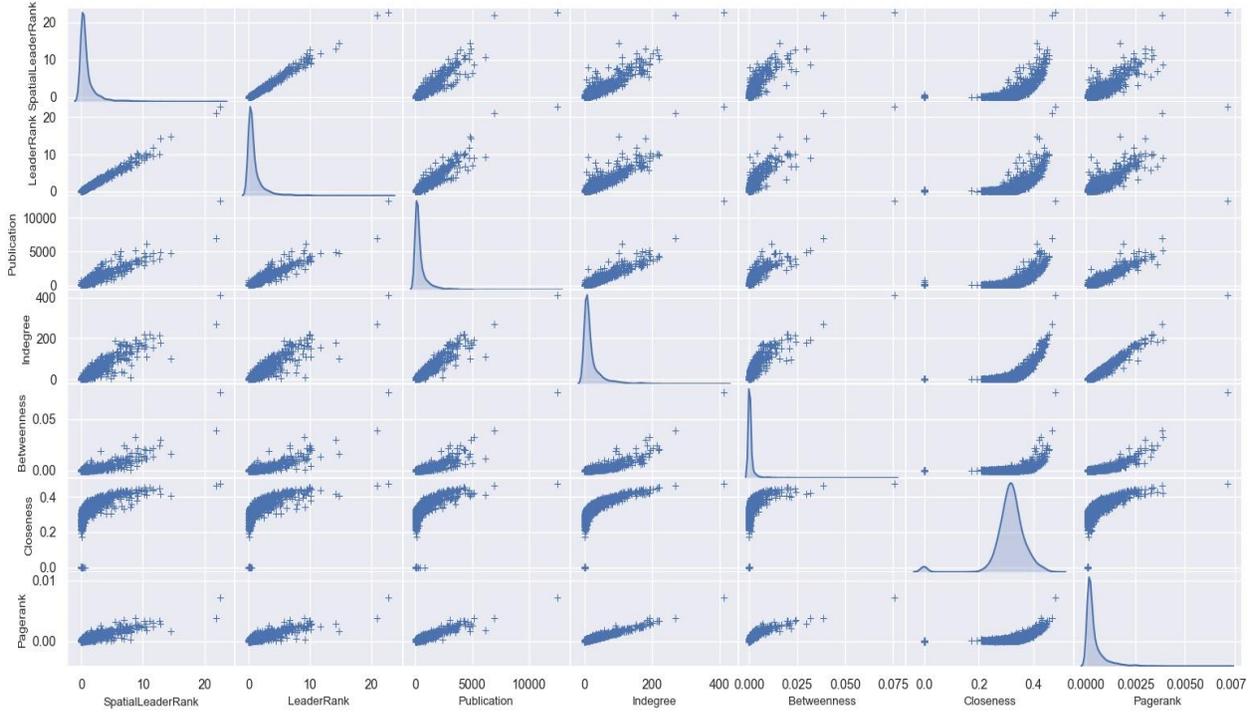

Figure 8. The pairwise relationships between SpatialLeaderRank and other indices.

Table 1. Spearman correlation coefficient between SpatialLeaderRank and other indices.

|  | SpatialLeaderRank | LeaderRank | Publication | Indegree | Betweenness | Closeness | PageRank |
| --- | --- | --- | --- | --- | --- | --- | --- |
| SpatialLeaderRank | 1.000 | 0.992 *** | 0.936 *** | 0.904*** | 0.875 *** | 0.858 *** | 0.870 *** |
| LeaderRank | 0.992 *** | 1.000 | 0.938 *** | 0.895 *** | 0.861 *** | 0.839 *** | 0.863 *** |
| Publication | 0.936 *** | 0.938 *** | 1.000 | 0.908*** | 0.918 *** | 0.833 *** | 0.915 *** |
| Indegree | 0.904 *** | 0.895 *** | 0.908 *** | 1.000 | 0.906 *** | 0.953 *** | 0.976 *** |
| Betweenness | 0.875 *** | 0.861 *** | 0.918 *** | 0.906 *** | 1.000 | 0.838 *** | 0.937 *** |
| Closeness | 0.858 *** | 0.839 *** | 0.833 *** | 0.953 *** | 0.838 *** | 1.000 | 0.904 *** |
| PageRank | 0.870 *** | 0.863 *** | 0.915*** | 0.976 *** | 0.937 *** | 0.904 *** | 1.000 |

*** $p<0.001$;

## 4.2.2. Evaluation

We evaluate the effectiveness of the SpatialLeaderRank by using it to predict institutions' academic impact, which is measured by four common indices. We also compare its performance with that of conventional indices. Citation count is a widely recognized measure of academic impact (Yan and Ding 2011). However, simple citation count is not robust against manipulations

(Hirsch 2005). To this end, Hirsch (2005) proposed h-index, the maximum value of h papers being cited at least h times for each entity (author, journal, institution, etc.). H-index combines the quantity and quality of publications and has been widely adopted by the scientific community (Lund 2019). Recently, Altmetrics indices[4] have emerged as popular tools to measure academic impact because they are less subject to the publication delays than citation. Similarly, the simple Altmetrics count is not robust against manipulations either. To this end, Askeridis (2018) proposed an Altmetrics-based H-index, namely, Mendeley-based H-index, which replaces the citation count with the view count of Mendeley readers. Mendeley is a free reference manager and academic social network. The view count of Mendeley users has been found to be associated with the later citation count (Aduku, Thelwall et al. 2017), because the view count is crowdsourced from the research community, and thus reflects the academic impact earlier than the citation count. In this study, we have two variants of the H-index, the citation-based index, and the Altmetrics-based H-index. Both of them wouldn't be large if an entity publishes either many publications with low citation/Altmetrics or very few publications with high citation/Altmetrics. In total, four evaluation metrics (citation, citation-based h-index, Altmetrics, and Altmetrics-based h-index) are employed as the model outcome to evaluate the effectiveness of SpatialLeaderRank.

We use the ROC curve analysis to examine the capability of SpatialLeaderRank in identifying top institutions with high academic impact (top 5% in terms of citation, citation-based h-index, Altmetrics, and Altmetrics-based h-index) and compare its performance with that of other conventional indices. ROC curve analysis is a graphic method to illustrate the performance of a binary classification system under different recognition thresholds (Hassan, Imran et al. 2017, Zhang, Wang et al. 2019). The larger AUC (area under the ROC curve) is, the better performance the focal binary classification system has. Figure 9 presents the ROC curve of the SpatialLeaderRank and other indices for identifying the top 5% institutions with high academic impact in terms of citation, citation-based h-index, Altmetrics, and Altmetrics-based h-index.

---

[4] We derive the altmetrics data from https://www.altmetric.com.

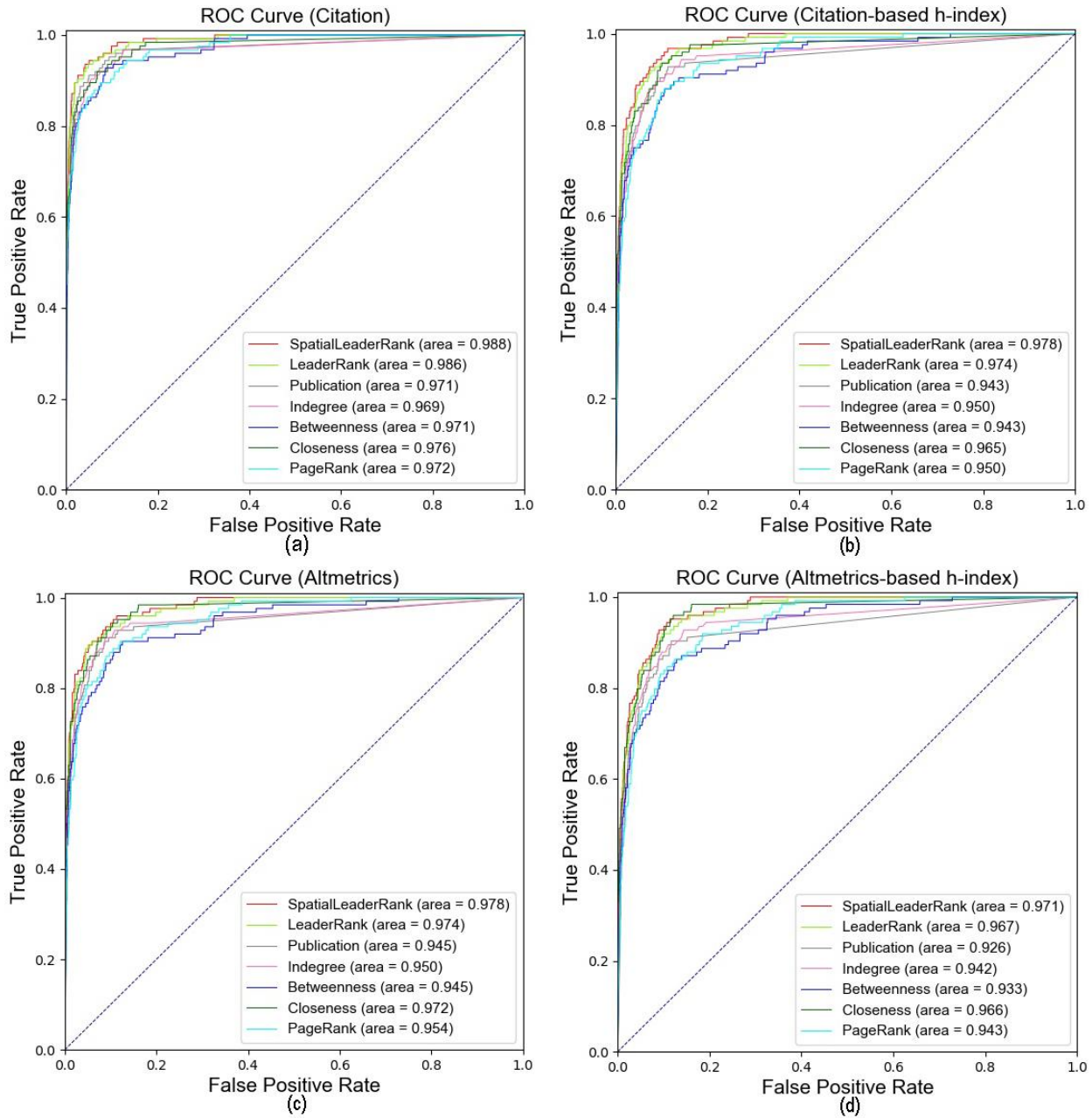

Figure 9 ROC curves for SpatialLeaderRank and other indices to identify top institutions with high academic impact.

As shown in Figure 9, all the indices have a reasonably high AUC (over 0.900) in terms of all the four evaluation metrics, indicating the generally good performance of all the indices in identifying the top institutions with high academic impact. Note that the high AUC value is expected because of the majority (95%) of institutions are classified as not influential. The proposed SpatialLeaderRank consistently outperforms other indices with all measures of academic impact. This indicates that integrating the spatial features does help to improve the capability in identifying top institutions with high academic impact. The larger SpatialLeaderRank score an institution has, the greater the leadership status that it has, and the higher academic impact that it generates. It's noteworthy that the AUC gaps between SpatialLeaderRank and other indices are more significant in h-index (citation-based/altmetrics-based) than that in citation/altmetrics counts. This is due to the fact that h-index (both the citation-based/altmetrics-based) is more robust than citation/ altmetrics, and can more closely reflect the academic impact. SpatialLeaderRank has more advantages in identifying top institutions with high academic impact compared with other conventional indices.

The ROC curve analysis only examines the overlap of top-k ranked institutions according to different indices, as it considers these top-k ranked institutions as an unordered set). Thus, it is meaningful to adopt another metric to further examine the relative ordering of the top-k ranked institutions. In this study, we adopt the *KSim* metric (Haveliwala 2003) which is based on Kendall's $\tau$ distance measure. Consider a ranking list by an index $\tau_1$ and a ranking list by an academic impact evaluation metric $\tau_2$. Let $U$ be the union of the institutions in $\tau_1$ and $\tau_2$. Let $\sigma_1$ be $U - \tau_1$ and $\tau_1'$ be the extension of $\tau_1$, where $\tau_1'$ contains $\sigma_1$ in addition to the existing ranked institutions in $\tau_1$. The rank of institutions in $\sigma_1$ is set to have the same ordinal rank in the end of $\tau_2$. Similarly, $\tau_2$ is extended to yield $\tau_2'$. The *Ksim* between $\tau_1$ and $\tau_2$ is defined as the Kendall's distance between $\tau_1'$ and $\tau_2'$, respectively:

$$KSim(\tau_1, \tau_2) = \frac{|(u,v): \tau_1', \tau_2' \text{ agree on order of } (u,v), u \neq v|}{(|U|)(|U|-1)}, \qquad (10)$$

In other words, *KSim* measures the probability of $\tau_1$ and $\tau_2$'s agreement on the relative ordering of a randomly selected pair of institutions $(u,v) \in U \times U$.

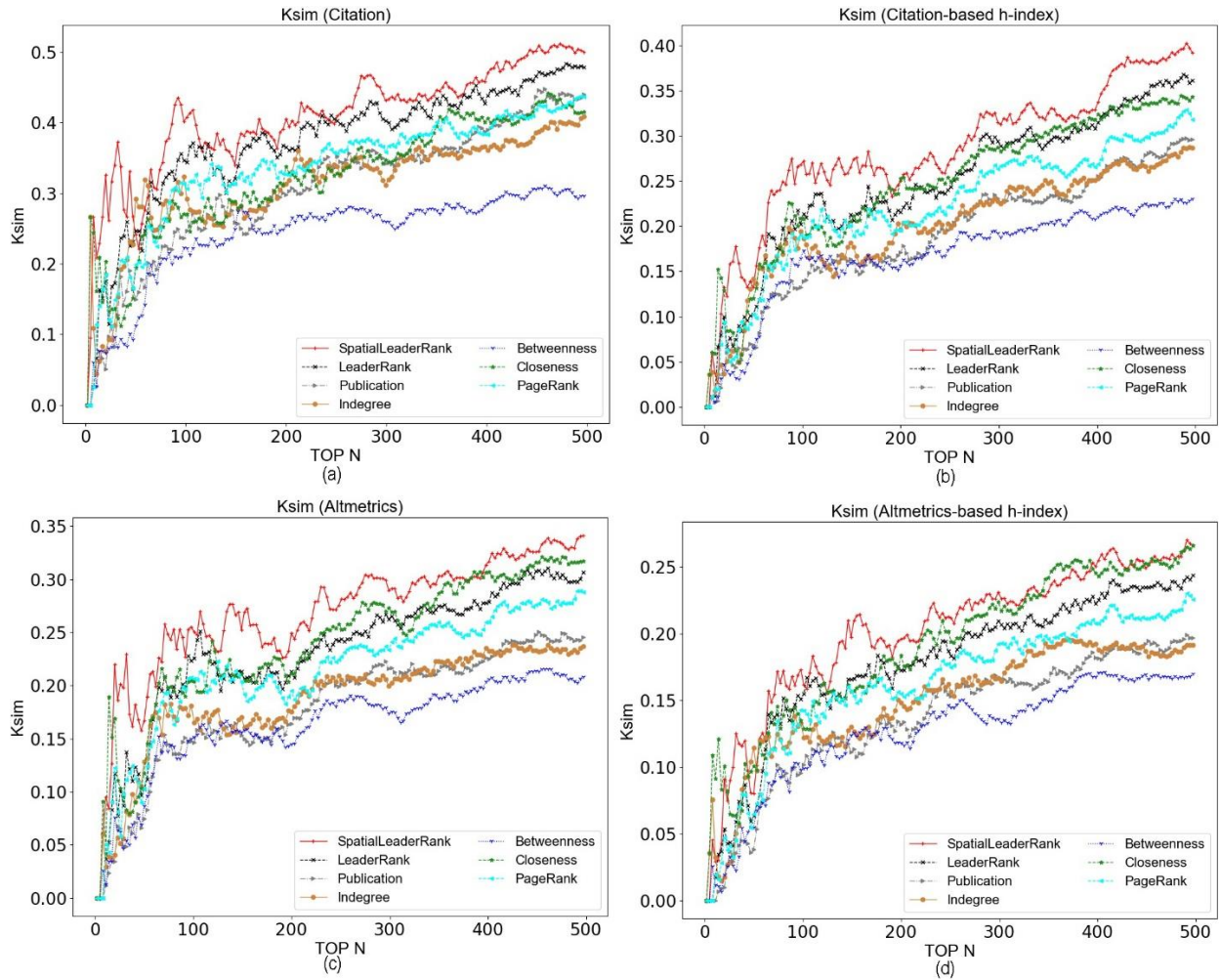

Figure 10. *KSim* between each index and four evaluation metrics from top 5 to top 500.

Figure 10 visualizes the *KSim* curve between each index and the four academic impact evaluation metrics from the top 5 institutions to the top 500 institutions. In general, SpatialLeaderRank consistently outperforms all other indices. In particular, when $N$ is small, the effect of extreme cases is more significant. In such a case, the advantage of SpatialLeaderRank over other indices is more obvious, indicating that SpatialLeaderRank is more robust to rare extreme cases. As $N$ increases, all indices' *Ksim* increase and become stable. Betweenness centrality has the worst performance according to *KSim* in all evaluation metrics. Betweenness centrality of an institution is proportional to the number of shortest paths traversing through the institution. The unsatisfactory performance of betweenness centrality is due to the fact that most of the top-ranked institutions by betweenness centrality are actually the local hubs that are connecting to local institutions. Publication count and indegree have similar performance. Interestingly, we find that if the publications and leading behaviors are mostly associated with local institutions, the academic impact does not get much improved. It's interesting that closeness centrality has a relatively high *Ksim* in altmetrics and particularly, altmetrics-based H-index. In altmetrics-based H-index, when *N* is larger than 300, closeness centrality's performance is close to the SpatialLeaderRank, primarily because institutions with high closeness centrality have shorter distance from other institutions, so the spread of research output is faster through the collaboration network.

## 4.2.3. Ranking institutions by SpatialLeaderRank, other conventional indices, and academic impact indices

We rank the institutions according to SpatialLeaderRank, the other conventional indices (LeaderRank, piublication, indegree, betweenness, closeness, and PageRank) and academic impact indices (citation, citation-based h-index, Altmetrics, and Altmetrics-based h-index). Table 2 presents the top 20 institutions. A larger table of top 100 institutions has been posted on our Github site[5]. Table 3 presents the rank of the SpatialLeaderRank-based top 20 institutions in other rankings. The top institutions ranked by SpatialLeaderRank are also highly ranked according to all the four academic impact indices. This strong association indicates that leading long-distance and cross-border collaborations generally lead to greater academic impact.

Four institutions (Harvard University, University of Oxford, University of Cambridge and University of California San Diego) are among the top 20 according to all indices. Particularly, Harvard University is ranked as the top one institution according to four academic impact indices, and top three according to the other indices. Chinese Academy of Sciences (CAS) and French National Centre for Scientific Research (French: Centre national de la recherche scientifique, CNRS) are two state research organizations in China and France, respectively. These two mega organizations have multiple research institutes, and have published many papers in the field (#1 for CAS and #7 for CNRS) (Table 2). It is not surprising to observe the high ranking of CAS and CNRS according to the collaboration indices. However, their performance in research impact is less significant than that of collaboration, indicating that CAS and CNRS are widely leading the research through collaborating with many other institutions, the academic impact still has room for further improvement. Similarly, we found that many other Chinese institutions are highly ranked in terms of the conventional indices, however, neither their SpatialLeaderRank nor academic impact is highly ranked. Their publications don't match their academic impact status, suggesting that Chinese institutions should focus on improving the quality of research instead of quantity only.

---

[5] https://github.com/chaochehe/Top-100-institutions

Table 2. Top 20 institutions' rank by SpatialLeaderRank and by other indices.

| Citation | Pagerank | Closeness | Betweenness | Indegree | Publication | LeaderRank | SpatialLeaderRank |
|---|---|---|---|---|---|---|---|
| Harvard Univ | Chinese Acad Sci | Chinese Acad Sci | Chinese Acad Sci | Chinese Acad Sci | Chinese Acad Sci | Chinese Acad Sci | Chinese Acad Sci |
| Chinese Acad Sci | Univ Sao Paulo | Harvard Univ | Harvard Univ | Harvard Univ | Harvard Univ | Harvard Univ | Harvard Univ |
| Univ Cambridge | Harvard Univ | UCSD | Univ Sao Paulo | Univ Michigan | Russian Acad Sci | CNRS | CNRS |
| Univ Oxford | CSIC | Univ Michigan | Univ Tokyo | UCSD | Univ Sao Paulo | Univ Tokyo | Univ Oxford |
| UCSD | UCSD | Univ Oxford | CSIC | Univ Oxford | SJTU | Univ Toronto | Univ Tokyo |
| UCSF | Univ Oxford | Univ Cambridge | Univ Oxford | Univ Cambridge | Univ Tokyo | Univ Copenhagen | Univ Cambridge |
| Stanford Univ | Univ Michigan | Univ Illinois | King Saud Univ | Univ Copenhagen | CNRS | CAMS | Univ Toronto |
| Univ Toronto | Zhejiang Univ | Stanford Univ | Univ Cambridge | Univ Illinois | Univ Toronto | Kyoto Univ | UCSD |
| MIT | Univ Cambridge | Univ Copenhagen | Kyoto Univ | Univ Sao Paulo | Seoul Natl Univ | UCSD | Univ Copenhagen |
| Univ Penn | Univ Tokyo | NCI | Univ Michigan | Zhejiang Univ | Zhejiang Univ | Univ Oxford | UCSF |
| CNRS | Univ Copenhagen | Univ Penn | Univ Copenhagen | CSIC | Fudan Univ | Univ Michigan | NCI |
| Univ Calif Berkeley | Peking Univ | Yale Univ | UCSD | Univ Tokyo | Univ Copenhagen | UCSF | Univ Michigan |
| Univ Copenhagen | Univ Illinois | Johns Hopkins Univ | Zhejiang Univ | Johns Hopkins Univ | UCSD | Johns Hopkins Univ | Univ Washington |
| Univ Michigan | Kyoto Univ | Univ Toronto | Univ Illinois | Peking Univ | Univ Michigan | Univ Cambridge | Johns Hopkins Univ |
| Univ Tokyo | SJTU | Ohio State Univ | Karolinska Inst | Univ N Carolina | Univ Oxford | Univ Penn | Stanford Univ |
| Univ Washington | Johns Hopkins Univ | Univ Wisconsin | CNRS | Stanford Univ | Peking Univ | Russian Acad Sci | Univ Penn |
| Duke Univ | UNAM | UCSF | Johns Hopkins Univ | Univ Maryland | Univ Illinois | Univ Sao Paulo | Duke Univ |
| UCLA | Karolinska Inst | Fudan Univ | NCI | SJTU | Univ N Carolina | Univ Washington | Kyoto Univ |
| Johns Hopkins Univ | Univ N Carolina | McGill Univ | UCL | Univ Pittsburgh | Sun Yat Sen Univ | Seoul Natl Univ | Univ Illinois |
| Yale Univ | Ohio State Univ | Univ Washington | Seoul Natl Univ | Univ Florida | Univ Cambridge | Osaka Univ | Univ N Carolina |

"CAMS": Chinese Academy of Medical Sciences; "CNRS": Centre national de la recherche scientifique (French National Centre for Scientific Research); "CSIC": Consejo Superior de Investigaciones Científicas (Spanish National Research Council); "HHMI": Howard Hughes Medical Institute; "MGH": Massachusetts General Hospital; "MIT": Massachusetts Institute of Technology; "NCI": National Cancer Institute, USA; "SJTU": Shanghai Jiao Tong University; "UCLA": University of California, Los Angeles; "UCSD": University of California San Diego; "UCSF": The University of California, San Francisco; "UCL": University College London; "UCLA": University of California, Los Angeles; "UNAM": National Autonomous University of Mexico;

| Altmetrics-based h-index | Altmetrics | Citation-based h-index |
| --- | --- | --- |
| Harvard Univ | Harvard Univ | Harvard Univ |
| Stanford Univ | Stanford Univ | MIT |
| MIT | MIT | UCSF |
| UCSF | Univ Cambridge | Stanford Univ |
| UCSD | UCSF | UCSD |
| Univ Calif Berkeley | UCSD | Univ Cambridge |
| Univ Cambridge | Univ Oxford | Univ Calif Berkeley |
| Univ Oxford | Univ Calif Berkeley | Univ Toronto |
| Yale Univ | Chinese Acad Sci | Univ Washington |
| HHMI | Univ Toronto | Univ Penn |
| Univ Penn | Yale Univ | MGH |
| Univ Toronto | Univ Penn | Yale Univ |
| MGH | Univ Washington | Univ Oxford |
| Univ Washington | Univ Copenhagen | Chinese Acad Sci |
| UCL | Univ Michigan | Johns Hopkins Univ |
| UCLA | UCL | HHMI |
| Columbia Univ | UCLA | Univ Copenhagen |
| Johns Hopkins Univ | Johns Hopkins Univ | Univ Michigan |
| Washington Univ | CNRS | Columbia Univ |
| Duke Univ | Duke Univ | CNRS |

Table 3. Comparison of institution rank by SpatialLeaderRank and other indices.

| Institution | SpatialLeaderRank | Citation | Citation-based h-index | Altmetrics | Altmetrics-based h-index | LeaderRank | Publication | Indegree | Betweenness | Closeness | PageRank |
|---|---|---|---|---|---|---|---|---|---|---|---|
| Chinese Acad Sci | 1 | 2 | 14 | 9 | 27 | 1 | 1 | 1 | 1 | 1 | 1 |
| Harvard Univ | 2 | 1 | 1 | 1 | 1 | 2 | 2 | 2 | 2 | 2 | 3 |
| CNRS | 3 | 11 | 20 | 19 | 26 | 3 | 7 | 81 | 16 | 64 | 73 |
| Univ Oxford | 4 | 4 | 13 | 7 | 8 | 10 | 15 | 5 | 6 | 5 | 6 |
| Univ Tokyo | 5 | 15 | 23 | 31 | 35 | 4 | 6 | 12 | 4 | 49 | 10 |
| Univ Cambridge | 6 | 3 | 6 | 4 | 7 | 14 | 20 | 6 | 8 | 6 | 9 |
| Univ Toronto | 7 | 8 | 8 | 10 | 12 | 5 | 8 | 32 | 23 | 14 | 31 |
| UCSD | 8 | 5 | 5 | 6 | 5 | 9 | 13 | 4 | 12 | 3 | 5 |
| Univ Copenhagen | 9 | 13 | 17 | 14 | 21 | 6 | 12 | 7 | 11 | 9 | 11 |
| UCSF | 10 | 6 | 3 | 5 | 4 | 12 | 26 | 29 | 30 | 17 | 32 |
| NCI | 11 | 27 | 35 | 30 | 34 | 21 | 49 | 21 | 18 | 10 | 26 |
| Univ Michigan | 12 | 14 | 18 | 15 | 25 | 11 | 14 | 3 | 10 | 4 | 7 |
| Univ Washington | 13 | 16 | 9 | 13 | 14 | 18 | 22 | 24 | 44 | 21 | 36 |
| Johns Hopkins Univ | 14 | 19 | 15 | 18 | 18 | 13 | 21 | 13 | 17 | 13 | 16 |
| Stanford Univ | 15 | 7 | 4 | 2 | 2 | 29 | 33 | 16 | 24 | 8 | 27 |
| Univ Penn | 16 | 10 | 10 | 12 | 11 | 15 | 24 | 25 | 32 | 11 | 28 |
| Duke Univ | 17 | 17 | 25 | 20 | 20 | 31 | 40 | 40 | 53 | 40 | 44 |
| Kyoto Univ | 18 | 35 | 57 | 77 | 93 | 8 | 23 | 35 | 9 | 69 | 14 |
| Univ Illinois | 19 | 24 | 44 | 37 | 56 | 22 | 17 | 8 | 14 | 7 | 13 |
| Univ N Carolina | 20 | 21 | 21 | 22 | 30 | 24 | 18 | 15 | 35 | 26 | 19 |

## 4.2.4. Results in Information Science & Library Science

To check the robustness of the study, we implement the same ROC curve analysis and Ksim measure analysis in ISLS field, as shown in Figure 11 and Figure 12. The results are consistent with those for pharmaceutical sciences. The proposed SpatialLeaderRank consistently outperformed other indices in predicting the academic impact of institutions. Similar to Section 4.2.3, we present the top 20 institutions in ISLS in Tables 4. A larger table of top 100 institutions is posted on Github site [6].

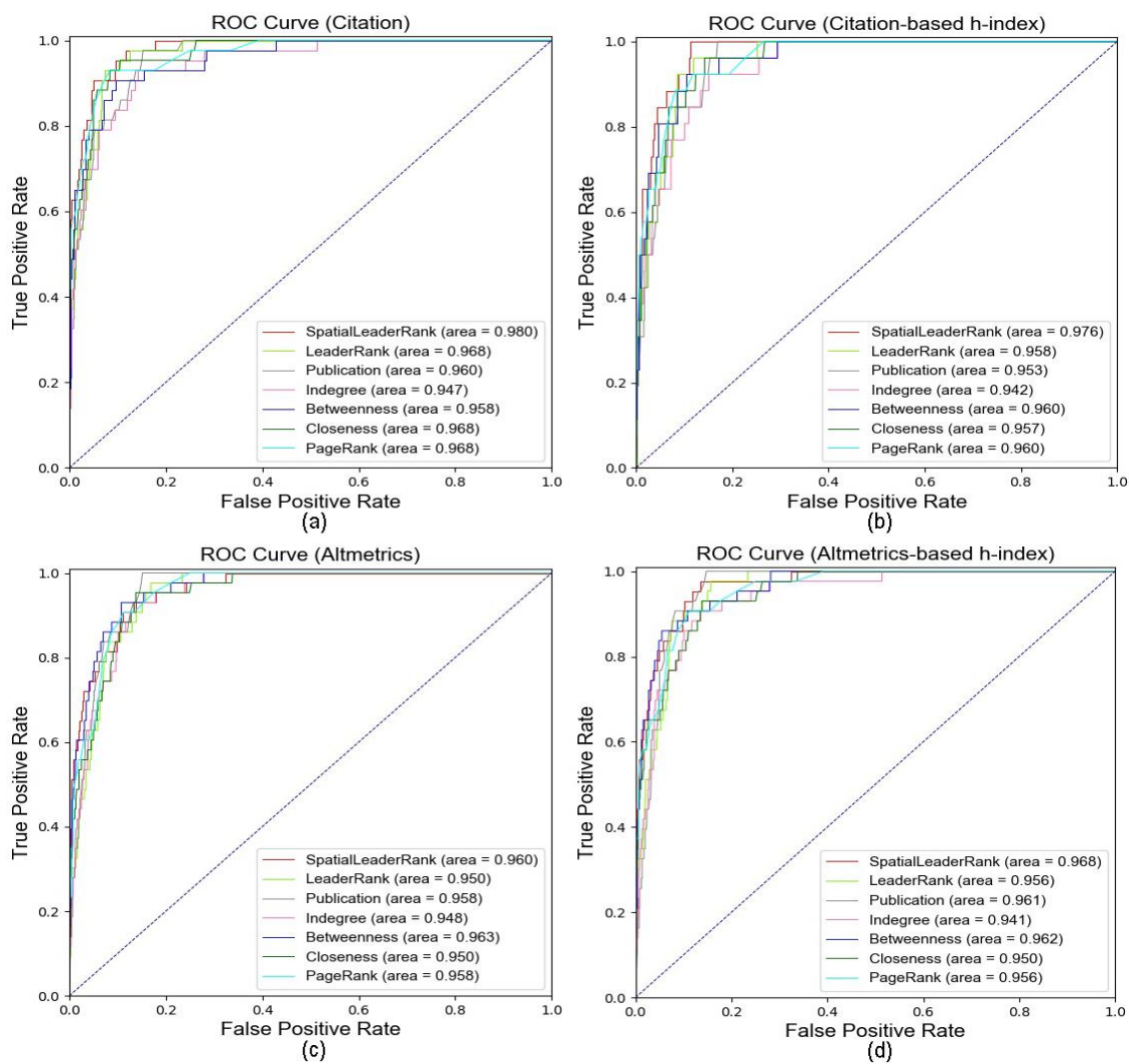

Figure 11 ROC curve for SpatialLeaderRank and other indices to identify top 5% institutions in ISLS.

---

[6] https://github.com/chaochehe/Top-100-institutions

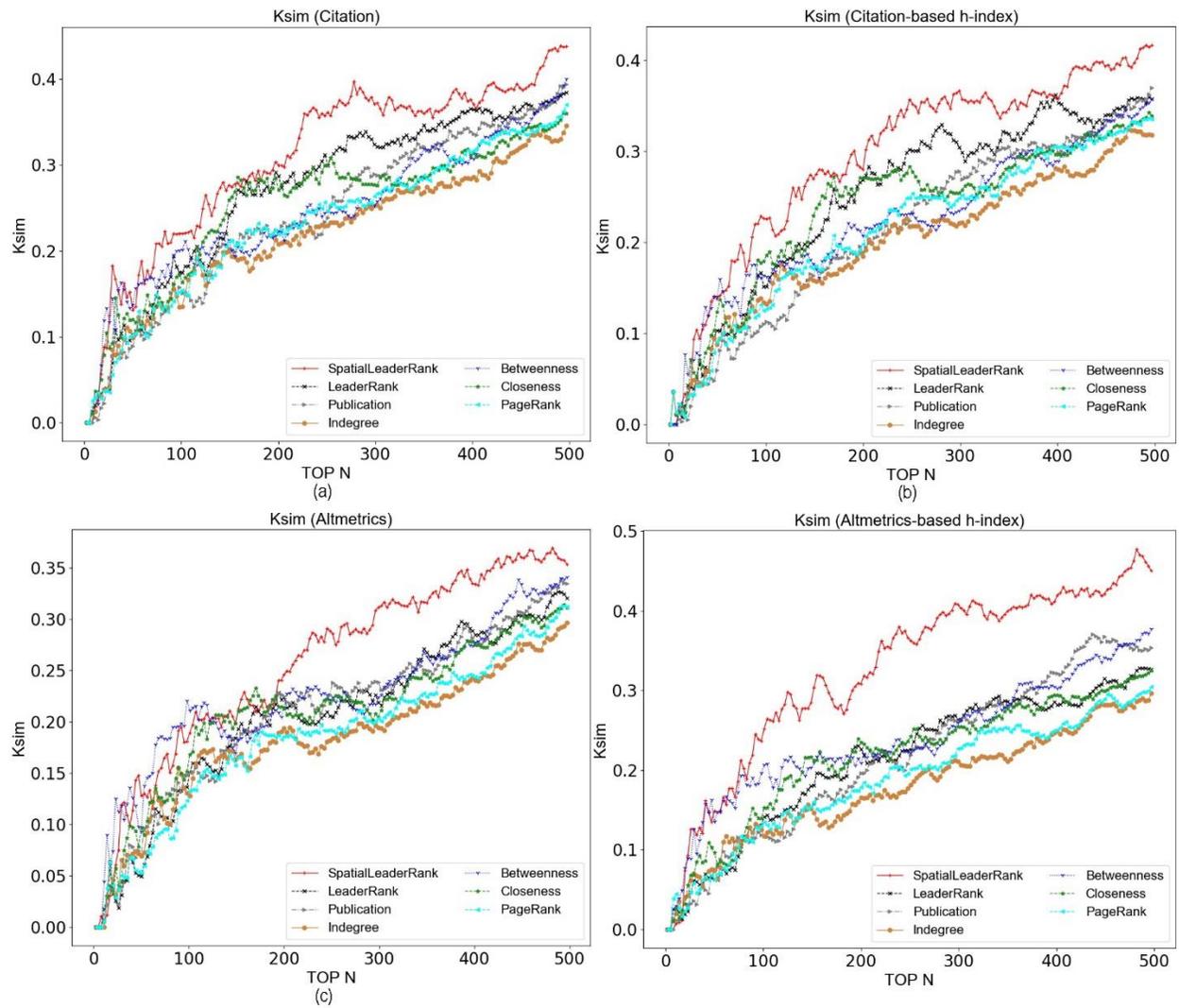

Figure 12 *KSim* between each index and four evaluation metrics from top 5 to top 500 in ISLS.

Table 4. Top 20 institutions' rank by SpatialLeaderRank and by other indices in ISLS.

| Pagerank | Closeness | Betweenness | Indegree | Publication | LeaderRank | SpatialLeaderRank |
|---|---|---|---|---|---|---|
| Indiana Univ | Wuhan Univ | Univ Wisconsin | Wuhan Univ | Wuhan Univ | Wuhan Univ | Univ Illinois |
| Univ Wisconsin | Indiana Univ | Univ Illinois | Indiana Univ | Univ Illinois | Univ Granada | City Univ Hong Kong |
| Wuhan Univ | Univ Wisconsin | Indiana Univ | Univ Wisconsin | Indiana Univ | City Univ Hong Kong | Univ Washington |
| City Univ Hong Kong | Nanyang Technol Univ | Univ Sheffield | Univ Granada | Univ Granada | Indiana Univ | Univ Wisconsin |
| Univ Illinois | City Univ Hong Kong | Wuhan Univ | City Univ Hong Kong | Nanyang Technol Univ | Univ Wisconsin | Univ Michigan |
| Univ Maryland | Univ Illinois | Univ Amsterdam | Univ Illinois | Univ Wisconsin | Chinese Acad Sci | Indiana Univ |
| Nanyang Technol Univ | Mcgill Univ | Univ Maryland | Beihang Univ | Univ Malaya | Univ Amsterdam | Univ Texas Austin |
| Arizona State Univ | Chinese Acad Sci | Univ N Carolina | Univ Carlos Iii Madrid | Russian Acad Sci | Univ Illinois | Georgia State Univ |
| Univ Arizona | Univ Arizona | Univ Carlos Iii Madrid | Nanyang Technol Univ | Univ N Carolina | Nanyang Technol Univ | Harvard Univ |
| Univ Amsterdam | Univ Amsterdam | Mcgill Univ | Chinese Acad Sci | Univ Maryland | Univ N Carolina | Univ N Carolina |
| Univ Texas Austin | Univ Sheffield | City Univ Hong Kong | Univ Maryland | Univ Carlos Iii Madrid | Mcgill Univ | Univ British Columbia |
| Chinese Acad Sci | Univ N Carolina | Univ Western Ontario | Univ Amsterdam | Univ Fed Minas Gerais | Arizona State Univ | Univ Maryland |
| Mcgill Univ | Arizona State Univ | Univ Granada | Natl Taiwan Univ | Univ Complutense Madrid | Univ Maryland | Natl Univ Singapore |
| Univ N Carolina | Univ British Columbia | Univ Texas Austin | Natl Univ Singapore | Univ Amsterdam | Natl Univ Singapore | Univ Pittsburgh |
| Univ Minnesota | Univ Maryland | Univ Washington | Univ N Carolina | Univ Sheffield | Zhejiang Univ | Nanyang Technol Univ |
| Univ Pittsburgh | Univ Hong Kong | Georgia State Univ | Yonsei Univ | Chinese Acad Sci | Nanjing Univ | Univ Amsterdam |
| Univ Michigan | Natl Univ Singapore | Natl Univ Singapore | Arizona State Univ | Mcgill Univ | Univ Arizona | Univ Manchester |
| Georgia State Univ | Univ Western Ontario | Univ Michigan | Univ Sheffield | Penn State Univ | Univ Minnesota | Georgia Inst Technol |
| Univ Carlos Iii Madrid | Univ Michigan | Chinese Acad Sci | Univ Minnesota | Natl Taiwan Univ | Univ Politecn Valencia | Mcgill Univ |
| Univ Hong Kong | Univ Minnesota | Nanyang Technol Univ | Drexel Univ | Univ Fed Santa Catarina | Max Planck Gesell | Wuhan Univ |

| Altmetrics-based h-index | Altmetrics | Citation-based h-index | Citation |
|---|---|---|---|
| Vanderbilt Univ | Univ Maryland | Univ Maryland | Univ Maryland |
| Harvard Univ | Wolverhampton Univ | Leiden Univ | Indiana Univ |
| Indiana Univ | Leiden Univ | Indiana Univ | Leiden Univ |
| Univ Maryland | Delft Univ Technol | Univ Amsterdam | Univ Arizona |
| Univ Michigan | Univ Sheffield | City Univ Hong Kong | Univ Arkansas |
| Wolverhampton Univ | Univ Michigan | Wolverhampton Univ | Wolverhampton Univ |
| City Univ Hong Kong | Indiana Univ | Vanderbilt Univ | Univ Amsterdam |
| Univ Amsterdam | Vanderbilt Univ | Harvard Univ | City Univ Hong Kong |
| Leiden Univ | Brunel Univ | Univ Arkansas | Vanderbilt Univ |
| Columbia Univ | Univ Amsterdam | Univ Wisconsin | Harvard Univ |
| Univ Wisconsin | Univ N Carolina | Temple Univ | Univ Wisconsin |
| Univ Washington | Univ Granada | Georgia State Univ | Temple Univ |
| Delft Univ Technol | Univ Wisconsin | Univ Arizona | Georgia State Univ |
| Univ Granada | City Univ Hong Kong | Univ Washington | Univ Illinois |
| Stanford Univ | Univ Illinois | Michigan State Univ | Wuhan Univ |
| Univ Sheffield | Univ Sydney | Univ Illinois | Univ Granada |
| Univ Sydney | Nanyang Technol Univ | Univ Granada | Univ British Columbia |
| Univ N Carolina | Univ Twente | Northwestern Univ | Univ N Carolina |
| Georgia State Univ | Univ Malaya | Univ Texas Austin | Univ Michigan |
| Univ Manchester | Harvard Univ | Univ N Carolina | Univ Washington |

# 5. Discussions and conclusions

In this paper, to address the spatial bias in research leadership flows in research collaboration, we examine the spatial distribution and the dynamic trend of research leadership flows in pharmaceutical sciences. We find that the distribution of research leadership flow distance presents a long-tail effect. In general, institutions are leading others at a growing distance. We observe that developing countries have been playing an increasingly important role in leading the research in pharmaceutical sciences.

Then, we construct a geographically-weighted and directed network, based on which we propose the SpatialLeaderRank. SpatialLeaderRank ranks the institutions integrating both topological features and spatial features. Comprehensive experiments with the data in both pharmaceutical sciences and ISLS demonstrate the superior performance of the proposed SpatialLeaderRank in predicting the academic impact of institutions. Leading institutions are identified and presented.

This study sheds light on the important association between long-distance and cross-border collaborations and academic impact. With the growing trend of cross-border collaborations, the distance between collaborators, particularly between the research leader and participators, should be an integral part to consider while examining the research leadership.

From a policy perspective, we found a clear rebalancing process between the research leadership mass in developed and developing countries. A number of eastern Asian countries, particularly China, is quickly emerging as a new global leader in pharmaceutical sciences. There are two main reasons for the change. First, the expenditure on research has been increasing rapidly in China (Basu, Foland et al. 2018). As of 2018, the research and development (R&D) expenditure in China was 1967.79 billion CNY[7], a 179% increase from 2010. China recently passed the European Union in R&D investment (Basu, Foland et al. 2018). Meanwhile, the research funding in the United States only increased by 57%[8]. Apparently, with abundant funding, Chinese institutions are playing the role as the leader more often. Second, the evaluation of research output in China is primarily based on quantitative measures such as publication number, the impact factor of journals, and citation count. The emphasis on these quantitative measures drove the whole academic community to publish as many papers as possible. Despite the success in publication number, the academic impact of Chinese institutions is still laid behind the Western institutions. It is suggested that policymakers in China shift the focus of the research evaluation towards the actual academic impact from quantitative measures. On 18/02/2020, the Ministry of Science and Technology and the Ministry of Education of China jointly published an announcement to urge the Chinese institutions to adopt a more scientific and influence-driven research evaluation approach[9]. This indicates the start of transformation from quantity to quality in the pharmaceutical sciences in China.

---

[7] http://www.stats.gov.cn
[8] https://www.statista.com

For the Western countries, including Europe and the United States, they are still playing the role as the major leaders in the field, more research expenditure is needed to maintain a good status. In general, cross-border collaboration is playing an increasingly important role in pharmaceutical research. Given the higher impact of long-distance collaborations, cross-border collaboration should be encouraged by means of joint-funding schemes and academic exchanges.

On the other hand, although our research suggests the long-distance collaboration is beneficial, it is not appropriate to directly take the spatial features in the evaluation of academic performance, as it is easy to manipulate. Long-distance and cross-border collaborations shall be encouraged to increase the chance to generate high impact research, but the impact of research should be purely based on its scientific merit.

The proposed SpatialLeaderRank is a general method that can be used to evaluate the leadership at the author and country levels. It is also applicable to examine other types of relationships, such as to evaluate the academic influence of scientific journals, to identify the key innovator in the industry by analyzing the patent-citation network and to identify the key moderator in the financial system by analyzing the guarantee-relationship network.

---

[9]http://www.moe.gov.cn/srcsite/A16/moe_784/202002/t20200223_423334.html